\documentclass[12pt,english]{article} 
\usepackage[T1]{fontenc}
\usepackage[latin1]{inputenc} 
\usepackage{doublespace} 
\usepackage{graphics} 
\usepackage{babel} 
\begin{document} 
\begin{center}

{\large MODIFIED DE BROGLIE-BOHM APPROACH TO QUANTUM MECHANICS} 
\vspace{.5 in}

Moncy V.  John

 {\it
Department of Physics, St.  Thomas College, Kozhencherri 689641, Kerala, India.}

\vspace{.5in}

{\bf Abstract} 
\end{center}

 A modified de Broglie-Bohm (dBB) approach to quantum mechanics is presented.  In
this new deterministic theory, which uses complex methods in an intermediate
step, the problem of zero velocity for bound states encountered in the dBB
formulation does not appear.  Also this approach is equivalent to standard
quantum mechanics when averages of observables like position, momentum and
energy are taken.


\medskip 
\noindent PACS No(s):  03.65.Bz

\newpage

\section{INTRODUCTION}

In the de Broglie-Bohm quantum theory of motion (dBB) [1-4], which provides a
deterministic theory of motion of quantum systems, the following assumptions are
made \cite{holland}:

									       (A1)
An individual physical system comprises a wave propagating in space and time,
together with a point particle which moves continuously under the guidance of
the wave.

									       (A2)
The wave is mathematically described by $\Psi (x,t)$, a solution to the
Schrodinger's wave equation.

									       (A3)
The particle motion is obtained as the solution $x(t)$ to the equation

\begin{equation} \dot{x} = \frac {\hbar }{2im} \frac {\left( \Psi ^{\star }\frac
{\partial \Psi }{\partial x}-\frac {\partial \Psi^{\star}} {\partial x}\Psi
\right)} {\Psi ^{\star} \Psi}, 
\label{eq:xdot1} 
\end{equation} 
with the initial
condition $x=x(0)$.  An ensemble of possible motions associated with the same
wave is generated by varying $x(0)$.

									       (A4)
The probability that a particle in the ensemble lies between the points $x$ and
$x+dx$ at time $t$ is given by $\Psi ^{\star }\Psi \; dx$.

\medskip

  In
spite of its success as a deterministic theory, this scheme has the difficulty
that, for bound state problems, the time-independent part $\psi$ of the wave
function is real and hence the velocity of the particle
turns out to be zero everywhere \cite{holland}.  This feature, that the particle
is always at rest, is not a satisfactory one in a quantum theory of motion.

To have a better understanding of this problem, let us recall that in the the
classical Hamilton-Jacobi theory, the trajectory for a particle may be obtained
from the equation

\begin{equation} 
\frac {\partial S}{\partial t} + H\left( x,\frac {\partial
S}{\partial x} \right) =0 
\end{equation} 
by first attempting to solve for the
Hamilton-Jacobi function $S=W-Et$ and then integrating the equation of motion

\begin{equation} p=m\dot{x} = \frac {\partial S}{\partial x} \label{eq:p}
\end{equation} 
with respect to time.  However, since the classical
Hamilton-Jacobi equation is not adequate to describe the micro world, one
resorts to the Schrodinger equation $i\hbar \partial \Psi/\partial t =
-(\hbar^{2}/2m)\partial^2 \Psi /\partial x^2 + V\Psi $ and solves for the wave
function $\Psi$.  The basic idea of dBB is to put $\Psi$ in the polar form
$Re^{iS/\hbar}$ and then to identify $S$ as the Hamilton-Jacobi function.  This
gives us Eq.  (\ref{eq:xdot1}), when the particle velocity is given by Eq.
(\ref{eq:p}).  For wave functions whose space part is real, the Hamilton's
characteristic function $W= constant$ and hence the velocity field is zero
everywhere, which is the difficulty mentioned above.  This is certainly an
undesirable feature in an ontological interpretation of quantum mechanics, which
claims to make the theory intelligible and causal \cite{bohmhiley}.  It is true
that dBB is altogether a non-Copenhagen physical theory which requires a
`quantum intuition' and it is not justifiable to criticize dBB for its failure
to return to classical mechanics.  However, the fact that the theory dubs the
material system in the most important and wide class of bound state problems as
stationary, irrespective of its position and energy, is counter-intuitive and
hence the theory cannot be termed as quantum theory of motion, at least in these
cases.

 Another deterministic approach to quantum mechanics, which also claims the
absence of this problem, is the Floyd-Faraggi-Matone (FFM) trajectory
representation developed by Floyd \cite{floyd}.  In one dimension, Floyd uses
the generalized Hamilton-Jacobi equation

\begin{equation} 
\frac {(W^{\prime})^2}{2m} +V-E=-\frac{\hbar^2}{4m}\left[
\frac{W^{\prime \prime \prime}}{W^{\prime}}-\frac{3}{2}\left( \frac {W^{\prime
\prime}}{W^{\prime}}\right)^2 \right], 
\end{equation} 
where $W^{\prime}$
($^{\prime}$ denotes $\frac{\partial }{\partial x}$) is the momentum conjugate
to $x$.  Recently Faraggi and Matone \cite{faraggi} have independently generated
the same equation (referred to as the quantum stationary Hamilton-Jacobi
equation) from an equivalence principle free from any axioms.  Floyd argues that
the conjugate momentum $W^{\prime}$ is not the mechanical momentum; instead,
$m\dot{x} = m \partial E/\partial W^{\prime}$.  The equation of motion in the
domain [$x,t$] is rendered by the Jacobi's theorem.  For stationarity, the
equation of motion for the trajectory time $t$, relative to its constant
coordinate $\tau$, is given as a function of $x$ by \cite{floyd,faraggi,carroll}

\begin{equation} t-\tau = \frac {\partial W}{\partial E},
\end{equation} 
where
$\tau $ specifies the epoch.  Thus the dBB and FFM trajectory representations
differ significantly in the use of the equation of motion, though they are based
on equivalent generalized Hamilton-Jacobi equations.  The FFM trajectory
representation does not claim equivalence with standard quantum mechanics in the
predictions of all observed phenomena.

									       In
this letter, we present a different and modified version of the dBB that
surmounts the problem mentioned earlier.  In addition,
its formulation is closer to the classical Hamilton-Jacobi theory than even the
conventional dBB.  We apply the scheme to a few simple problems and find that it
is capable of providing a deeper insight into the quantum phenomena.  Also the
new scheme is constructed in such a way that it is equivalent to standard
quantum mechanics when averages of observables like position, momentum and
energy are taken.

In an intermediate step in the new formulation, we
consider the analytic continuation of the spacetime into complex, producing a
complex spacetime.  It is well known that such complex spacetimes are put to
remarkable use in the theory of relativity.  For example, the transformation of
one solution of Einstein equation into another can on many occasions be
accomplished by means of a simple complex switch of coordinates.  Open and
closed Friedmann models, de Sitter and ant-de Sitter spacetimes, Kerr and
Schwarschild metrics etc.  are related by complex substitutions.  The use of
complex variables in relativity extends to more sophisticated ones like spin
coefficient formalism, Ashtekar formalism, Twistor theory etc.  Recently, it was
demonstrated that a Friedmann model with complex scale factor leads to a
remarkably good nonsingular cosmological model, with none of the main
cosmological problems in standard cosmology appearing in it \cite{mvj}.  The
Wick rotation of the time coordinate $t$ in Minkowski space to $i\;t$, which
changes its metric into Euclidean or reversing, going from Euclidean to
Minkowski metric is already familiar in quantum field theory and quantum gravity
as Euclidean Feynman path integration.  Even in quantum mechanics, the technique
of considering the space variable as complex is familiar \cite{landau}.  Some
more specific usage of complex methods include the interpretation of spin
angular momentum as orbital momentum but taken about a complex center of mass
and the interpretation of magnetic moment as the electric dipole moment taken
about a complex center of charge \cite{newmann}.  The present modified dBB
scheme may be considered as yet another application of complex methods in
quantum theory.

									       The
paper is organized as follows.  In Sec. 2, we present the basic formalism and
in Sec. 3, we show that the new scheme is
constructed to give the same statistical predictions as that in standard quantum
mechanics.  In section 4, we apply the scheme to a few simple problems while
section 5 deals with the problem of action variable in the new scheme.  The last
section gives our conclusions and discussion.

\section{THE GENERAL FORMALISM}

We first note that in the dBB, the substitution $\Psi=Re^{iS/\hbar}$ in the
Schrodinger equation gives rise to two coupled partial differential equations,
one of which is a conservation equation for probability density, and this leads
to a situation apparently different from the Hamilton-Jacobi theory for
individual particles.  On the other hand, we note that a change of variable
$\Psi \equiv {\cal N}e^{i\hat{S}/\hbar}$ in the Schrodinger equation (where
${\cal N}$ is a normalisation constant) gives a quantum-mechanical
Hamilton-Jacobi equation \cite{goldstein}

\begin{equation} \frac {\partial \hat{S}}{\partial t} + \left[
\frac{1}{2m}\left( \frac {\partial \hat{S}}{\partial x}\right)^2 +V\right] =
\frac{i\hbar}{2m} \frac{\partial^2 \hat{S}}{\partial x^2},
 \end{equation}
 which
is closer to its classical counterpart and which also has the correct classical
limit.  Considering $\hat{S}$ as a modified Hamilton-Jacobi function brings the
expression (\ref{eq:p}) for the conjugate momentum to the form

\begin{equation} m\dot{x} \equiv \frac {\partial \hat{S}}{\partial x}= \frac
{\hbar }{i} \frac {1}{\Psi}\frac {\partial \Psi}{\partial x}.  \label{eq:xdot2}
\end{equation}
In this paper, we attempt to modify the dBB scheme in a manner
similar to the latter approach; i.e., we retain assumptions A1 and A2 whereas in
A3, we identify $\Psi \equiv {\cal N}e^{i\hat{S}/\hbar}$ and use the expression
(\ref{eq:xdot2}) as the equation of motion.

									       Note
that in this case, with $\Psi$ as the standard solution to the Schrodinger
equation in any given problem, $\dot{x}$ turns out to be a complex variable.
Consequently, also $x$ should be complex.  As
mentioned in the introduction, considering $x$ as a complex variable is not
entirely new, even in quantum mechanics.  While discussing the quasi-classical
problems, Landau and Lifshitz \cite{landau} use the technique of `complex
classical paths' extensively, as an intermediate step.  We propose to consider
$x$ as complex in the general formalism of quantum mechanics, in the context of
dBB theory.  It is clear that when $x$ is a complex variable, the Schrodinger
equation itself becomes a complex differential equation.  In the following, we
consider only those solutions to such equation, obtained by replacing the real
$x$ with $x\equiv x_r + i x_i$ in the standard solution.  From here onwards, we
consider $x$ as a complex variable.

We restrict ourselves to single particles in one dimension for simplicity.  The
formal procedure adopted is as follows:  Identifying the standard solution $\Psi
$ as equivalent to $ {\cal N}e^{i\hat{S}/\hbar}$, Eq.(\ref{eq:xdot2}) helps us
to obtain the expression for the velocity field $\dot{x}$ in the new scheme.
This, in turn, is integrated with respect to time to obtain the complex path
$x(t)$, with an initial value
$x(0)$.  This complex solution is obviously quite different from the
corresponding dBB solution.  The connection with the real physical world is
established by postulating that the physical coordinate or trajectory of the
particle is the real part $x_r(t)$ of its complex path $x(t)$.  (It is worth
recalling that this procedure is commonly adopted in physical problems which
involve differential equations with complex solutions, as in the case of a
classical harmonic oscillator.)  It can be seen that in the general case, also
this modified dBB trajectory $x_r(t)$ is different from the dBB trajectory.

 The real part of the velocity field $\dot{x}_r$, at any point $x$ in the
complex plane may be found from Eq.  (\ref{eq:xdot2}) as

\begin{equation} \dot{x}_r = \frac {\hbar }{2im} \frac {\left[ \Psi ^{\star
}\frac {\partial \Psi }{\partial x}-(\frac {\partial \Psi} {\partial
x})^{\star}\Psi \right]} {\Psi ^{\star} \Psi}, \label{eq:xdot4} 
\end{equation}
an expression quite analogous to the expression for the velocity field in the
dBB scheme (but not identical with it, since $\dot{x}_r$ is now defined over the
entire $x$-plane).  One can also write $\dot{x}_r$ as

\begin{equation} \dot{x}_r = \frac {\hbar }{2im} \frac {\left[ \Psi ^{\star
}\frac {\partial \Psi }{\partial x_r}-(\frac {\partial \Psi} {\partial
x_r})^{\star}\Psi \right]} {\Psi ^{\star} \Psi}, \label{eq:xdot5}
\end{equation}
where use is made of the fact that for complex derivatives, $\frac {\partial
\Psi }{\partial x}=\frac {\partial \Psi }{\partial x_r}$.  It is easy to see
that this is identical to the dBB velocity field at all points along the real
axis $x=x_r$.

We shall note that the new scheme differs from the dBB also in one important
aspect.  Here, when the particle corresponds to some point $x$ in the complex
plane, its physical coordinate is $x_r$ and the physical velocity is
$\dot{x}_r$, evaluated at $x$ using (\ref{eq:xdot4}).  Thus for the same
physical coordinate $x_r$, the particle can have different physical velocities,
depending on the point $x$ through which its complex path passes.  This in
principle rules out the possibility of ascribing simultaneously well-defined
physical position and velocity variables for the particle.  The situation here
is very different from the dBB in which for given physical position of the
particle, also the velocity is known, as obtainable from Eq.  (\ref{eq:xdot1}).

	Let us
consider the case in
which $\hat{S}=\hat{W}-Et$,
where $E$ and $t$ are assumed to be real.  Then the Schrodinger equation gives
us an expression for the energy of the particle

\begin{equation} E=\frac{1}{2} m\dot{x}^2+V(x)+\frac{\hbar}{2i} \frac{\partial
\dot{x}}{\partial x}.  
\label{eq:E} 
\end{equation} 
The last term resembles the
quantum potential in the dBB theory.

\section{EXPECTATION VALUES}

We recall that in this
scheme, for
the same quantum
state the particle can have any number of complex paths, depending on the
initial value $x(0)$.  When we calculate the expectation values of dynamical
variables, it should be such that this is not done for a single particle which
belongs to a particular complex path, but over an ensemble of particles which
corresponds to all possible complex paths.  We modify assumption (A4) by
postulating that the average of an observable $O$ is obtained using the measure
$\Psi^{\star} \Psi $ as

\begin{equation} <O> = \int_{-\infty}^{\infty} O \Psi^{\star} \Psi \; dx,
\end{equation} 
where the integral is taken along the real axis.  This can be
explicitly written for $x$, $p$ and $E$ as

\begin{equation}
 <x> = \int x \Psi^{\star} \Psi \; dx , 
\end{equation}

\begin{equation}
 <p> = \int p \Psi^{\star} \Psi \; dx =\int \Psi^{\star}
\frac{\hbar}{i} \frac{\partial \Psi}{\partial x} \; dx \label{eq:pexp}
\end{equation} 
and

\begin{equation} 
<E> = \int E \Psi^{\star} \Psi \; dx =\int \Psi^{\star} \left(
\frac{-\hbar^2}{2m} \frac{\partial^2 }{\partial x^2} +V(x)\right) \Psi \; dx ,
\label{eq:Eexp} 
\end{equation} 
where use is made of equations (\ref{eq:xdot2})
and (\ref{eq:E}) in equations (\ref{eq:pexp}) and (\ref{eq:Eexp}), respectively.
Note that we did not need to make the conventional operator replacements in the
above expressions.  Since the integral is taken along the real axis, these
coincide with the corresponding quantum mechanical expectation values.  Thus the
new scheme is equivalent to standard quantum mechanics when averages of
observables like position, momentum and energy are computed using the above
prescription.

\section{SIMPLE APPLICATIONS}

\begin{figure}[ht] 
\centering{\resizebox {1.05 \textwidth} {0.7 \textheight }
{\includegraphics {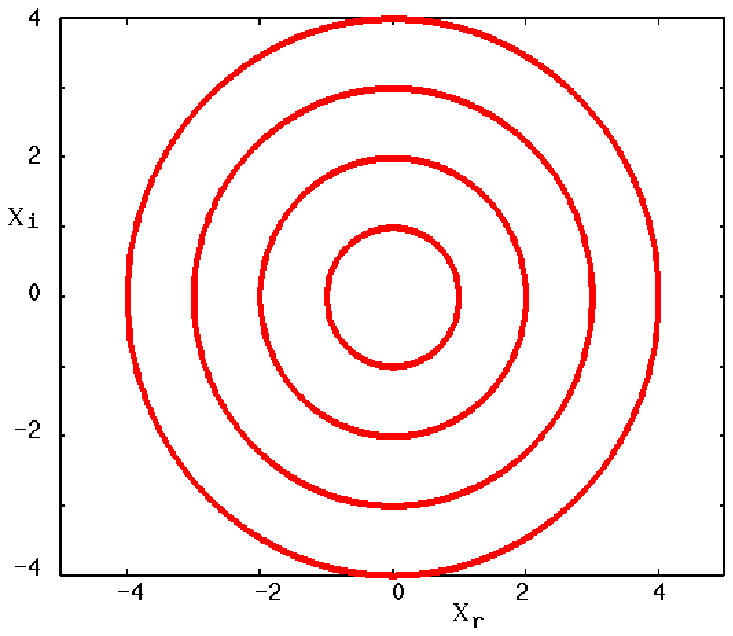}} 
\caption{Complex paths in the $X$-plane ($X\equiv
\alpha x$) for the $n=0$ harmonic oscillator case, where contours are plotted
for $X(0)$ =1, 2, 3 and 4.  } \label{fig:shm0}} 
\end{figure}

To see how the scheme works, let us consider the
example of the ground state solution $\Psi_0$ of the Schrodinger equation for
the harmonic oscillator in one dimension, the space part of which is real.  We
have

\begin{equation} 
\Psi_0 ={\cal N}_0 e^{-\alpha^{2}x^{2}/2} e^{-iE_0t/\hbar}.
\end{equation} 
The velocity field in the new scheme is given by Eq.
(\ref{eq:xdot2}) as

\begin{equation} 
\dot{x}= -\frac{\hbar}{im} \alpha^{2} x,\label{eq:xdotshm0}
\end{equation} 
whose solution is

\begin{equation} 
x=A e^{i\hbar \alpha^{2} t/m }.  
\end{equation} 
This is an
equation for a circle of radius $|A|=|x(0)|$ in the complex plane (Fig.
\ref{fig:shm0}).  As proposed earlier, we take the real part of this expression,

\begin{equation} 
x_r=A \cos(\hbar \alpha^2t/m),
 \end{equation} 
 [where it is
chosen that at $t=0$, $x(0)\equiv A$ is real] as the physical coordinate of the
particle.  It shall be noted that this is the same classical solution for a
harmonic oscillator of frequency $\omega_0 = \hbar \alpha^2 /m$.  However, the
nonclassical feature here is that we have a trajectory for every initial value
$x(0)$, though the energy $E$ has the value $\frac{1}{2}\hbar \omega_0$ for all
such trajectories.

\begin{figure}[ht] 
\centering{\resizebox {1.05\textwidth} {0.7 \textheight }
{\includegraphics {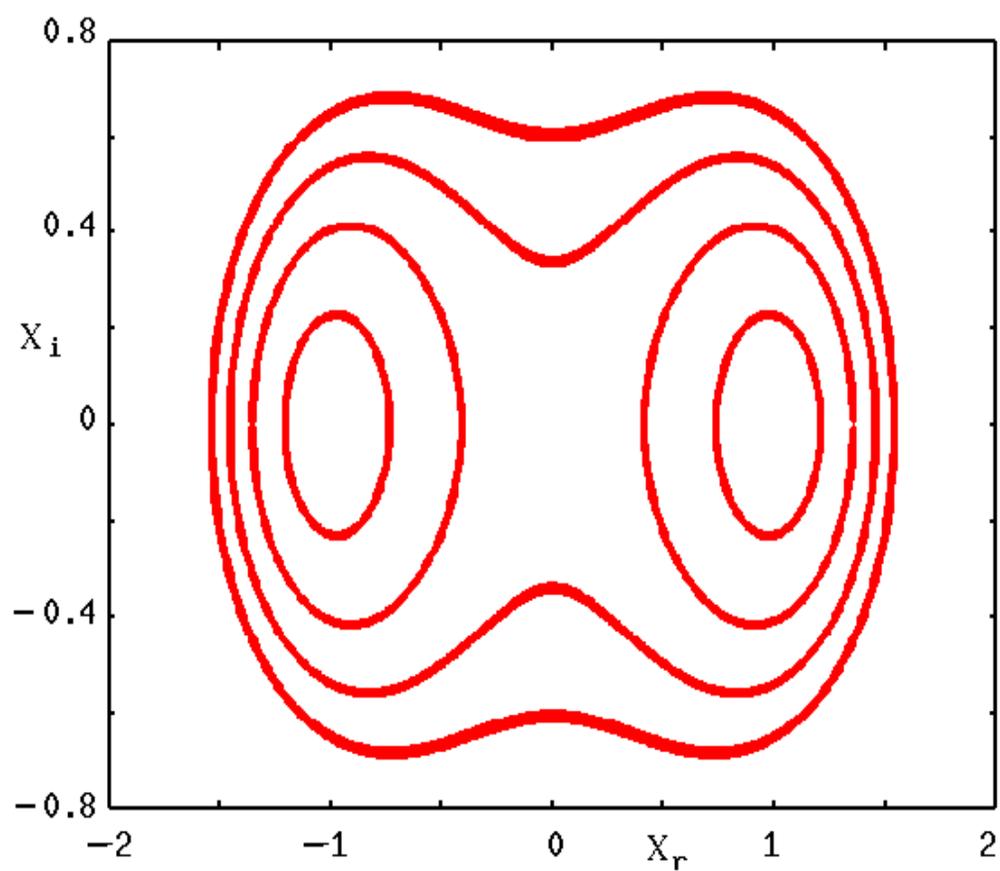}} \caption{Complex paths in the $X$-plane ($X\equiv
\alpha x$) for the $n=1$ harmonic oscillator case, where contours are plotted
for $X(0)$ =1.2, 1.35, 1.45 and 1.55.  } \label{fig:shm1}} 
\end{figure}

\begin{figure}[ht] 
\centering{\resizebox {1.05\textwidth} {0.7 \textheight }
{\includegraphics {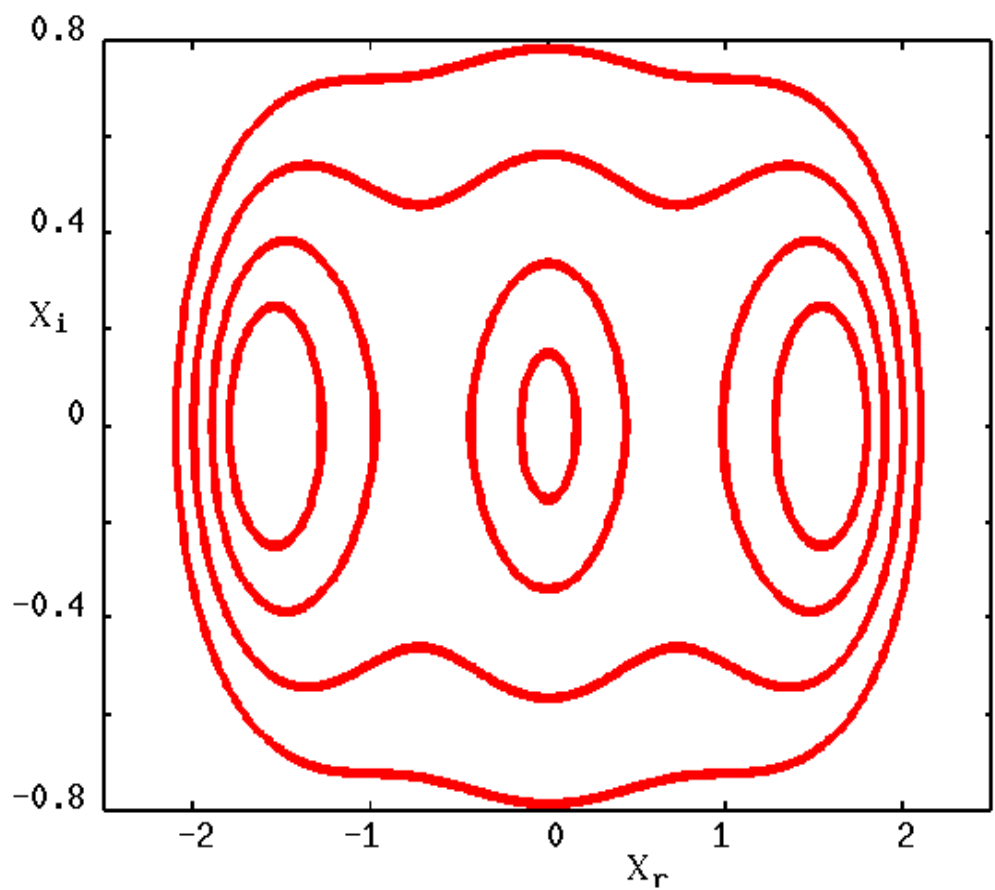}} \caption{Complex paths in the $X$-plane ($X\equiv
\alpha x$) for the $n=2$ harmonic oscillator case, where contours are plotted
for $X(0)$=1.8, 1.9, 2.0 and 2.1.}  \label{fig:shm2}} 
\end{figure}

									       We
can adopt this procedure to obtain the velocity field, also for higher values of
the quantum number n.  For $n=1$, we have

\begin{equation} 
\Psi_1 = {\cal N}_1 e^{-\alpha^{2}x^{2}/2}\; 2\alpha x \;
e^{-iE_1 t/\hbar}, 
\end{equation} 
from which

\begin{equation} 
\dot{x}=\frac{\hbar}{im} \left(-\alpha^{2} x +
\frac{1}{x}\right).
\label{eq:xdotshm1} 
\end{equation} The solution to this
equation can be written as

\begin{equation} (\alpha x -1)(\alpha x +1)=A e^{2i\hbar\alpha ^{2} t/m}
\end{equation} 
or

\begin{equation} x=\frac {1}{\alpha }\sqrt {1+A e^{2i\hbar\alpha ^{2} t/m}}.
\end{equation} 
Here, the solution is a product of two circles centered about
$\alpha x=\pm 1$, which is plotted in Fig.  \ref{fig:shm1}.  The physical
coordinate of the particle is again given by the real part of this expression.
For $n=2$, the solution can similarly be constructed as

\begin{equation} 
4\alpha x \left( \alpha x -\sqrt{5/2}\right)^{2}\left(\alpha x
+\sqrt{5/2}\right)^{2}=A e^{5i\hbar\alpha ^{2} t/m}.  
\end{equation} 
The complex
path in the $x$-plane is plotted in Fig.  \ref{fig:shm2}.  For cases with $n\geq
1$, these paths deviate from the classically expected circular ones.  Note that
in all these cases, the velocity fields are not zero everywhere, contrary to
what happens in the dBB.

									       Now
let us apply the procedure to some other stationary states which have complex
wave functions.  For a plane wave, we have

\begin{equation} 
\Psi =A e^{ikx}e^{-iEt/\hbar}, 
\end{equation} 
so that the
equation of motion is $\dot{x}=\hbar k/m$ and this gives

\begin{equation}
 x=x(0)+\frac{\hbar k}{m} t, 
 \end{equation} 
 the classical
solution for a free particle.

\begin{figure}[ht] 
\centering{\resizebox {1.05\textwidth} {0.7\textheight }
{\includegraphics {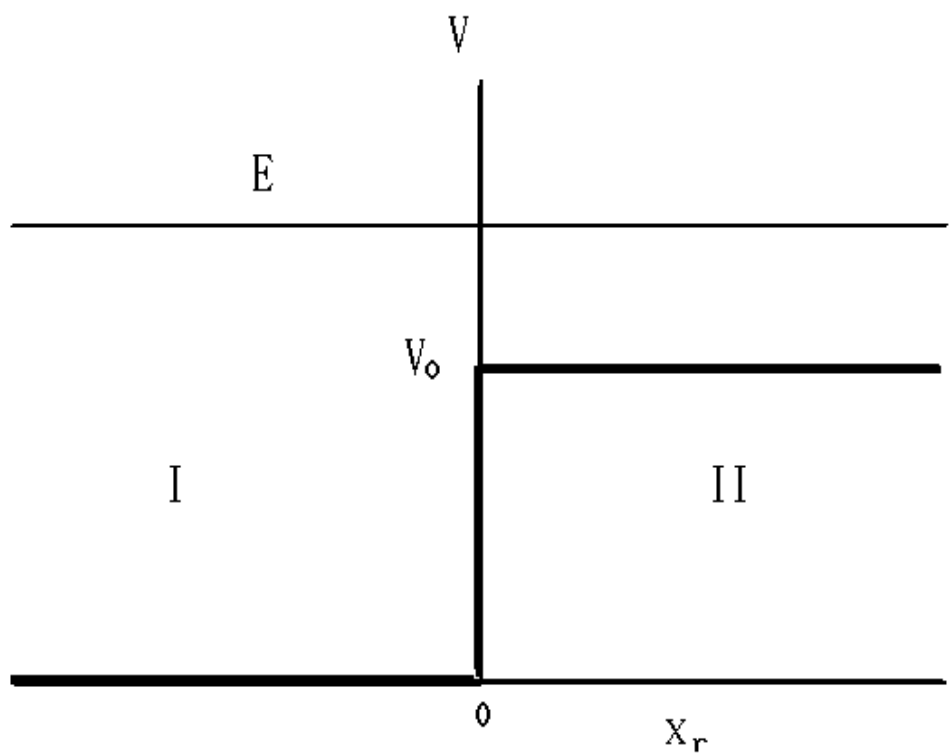}} 
\caption{ Potential step, with $V=0$ for
$x_r<0$, $V$ = $V_0$ for $x_r >0$ and energy $E>V_0$.  } \label{fig:step0}}
\end{figure}

\begin{figure}[ht] 
\centering{\resizebox {1.05 \textwidth} {0.7\textheight }
{\includegraphics {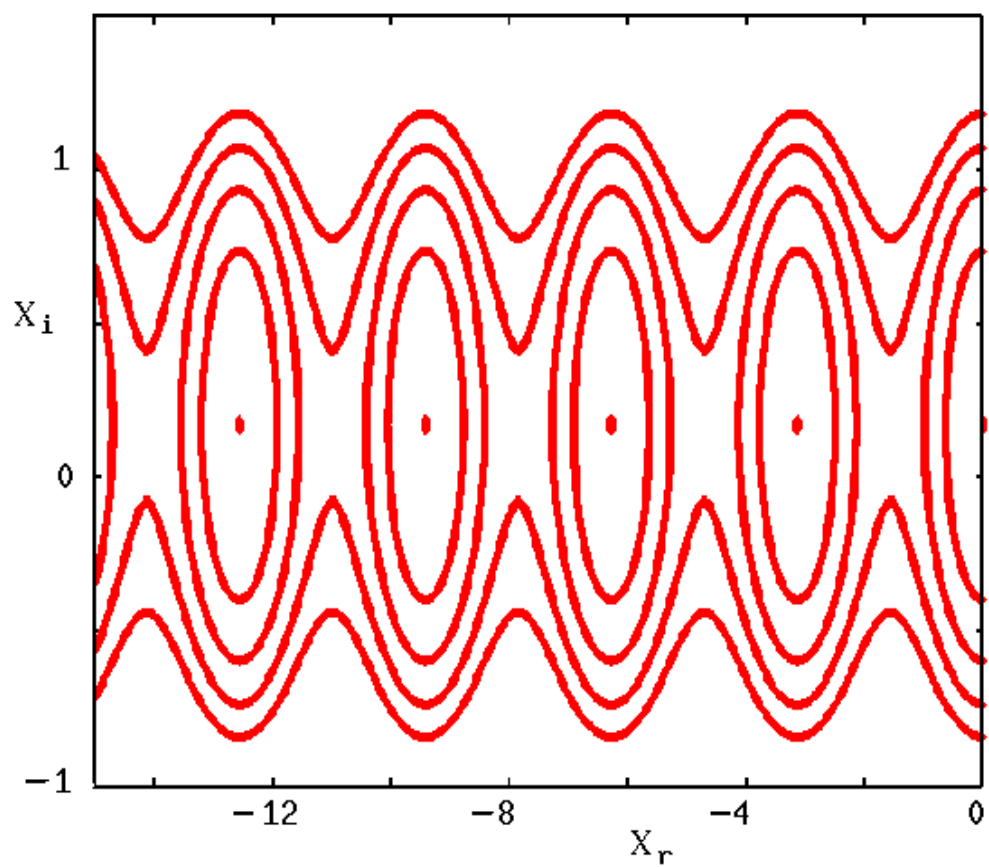}} \caption{ Complex paths for particles
approaching the potential step.  Contours for $c$ = -4, -3, -2, -1 and 0 are
plotted.  } \label{fig:step1}} 
\end{figure}

 As
another example, consider a particle with energy $E$ approaching a potential
step with height $V_0$, shown in Fig.  \ref{fig:step0}.  In region I, we have

\begin{equation} 
\psi_I = e^{ikx}+Re^{-ikx}, \qquad k=\sqrt{2mE/\hbar^2}
\end{equation} 
and, in region II,

\begin{equation} 
\psi_{II} = Te^{iqx}, \qquad q=\sqrt{2m(E-V_0 )/\hbar^2}.
\end{equation} 
The velocity fields in the two regions are given by

\begin{equation} 
\dot{x}_I = \frac {\hbar k}{m}\left(\frac{e^{ikx}-Re^{-ikx}
}{e^{ikx}+Re^{-ikx}}\right) 
\end{equation} 
and

\begin{equation} \dot{x}_{II} =\frac{\hbar q}{m}, 
\end{equation}
 respectively.
The contours in the complex $x$-plane in region I, for a typical value of the
reflection coefficient $r\equiv R^2=1/2$, are given by

\begin{equation} 
\sqrt{2} \cos 2kx_{Ir}-e^{-2kx_{Ii}}-\frac{1}{2}e^{2kx_{Ii}}=c
\end{equation} 
(where $x_{Ir}$ and $x_{Ii}$ are, respectively, the real and
imaginary parts of $x_I$), and are plotted in Fig.  \ref{fig:step1}.  Note that
also this case is significantly different from the corresponding dBB solution.

 Lastly,
we consider a nonstationary wave function, which is a spreading wave packet.
Here, let the propagation constant $k$ has a Gaussian spectrum with a width
$\Delta k \sim 1/\sigma $ about a mean value
$\bar{k}$. 
 The wave function is given by

\begin{equation} 
\Psi (x,t)={\cal N} \left( \frac {2\pi \sigma }{\sigma^2
+i\hbar t/m}\right)^{1/2} \exp \left[ - \frac {(x-i\sigma
^2\bar{k})^2}{2(\sigma^2 + i\hbar t/m)} - \frac { \sigma^2 + \bar{k}^2}{2}
\right].  
\end{equation} 
The velocity field is obtained from (\ref{eq:xdot2}) as

\begin{equation} 
\dot{x} = - \frac {\hbar}{im} \left( \frac {x-i\sigma ^2
\bar{k}}{\sigma^2+i\hbar t/m}\right) 
\end{equation} 
Integrating this expression,
we get

\begin{equation} 
x(t)= x(0)+ \frac{\hbar \bar{k}}{m} t+ i \frac{\hbar}{m}
\frac{x(0)}{\sigma^2}t. 
 \end{equation} 
 Separating the real and imaginary parts
of this equation (and assuming $\bar{k}$ is real), we get,

\begin{equation} 
x_r(t)= x_r (0)+ \frac{\hbar \bar{k}}{m} t+ \frac{\hbar}{m}
\frac{x_i (0)}{\sigma^2}t, 
\end{equation}

\begin{equation} 
x_i (t)= x_i (0)+ \frac{\hbar}{m} \frac{x_r (0)}{\sigma^2}t .
\end{equation} 
For the particle with $x(0)=0$, we obtain $x_r(t) =(\hbar
\bar{k}/m)t$ and $x_i(t)=0$; i.e., this particle remains at the center of the
wave packet.  Other particles assume different values for $x(t)$ at time $t$ as
given by the above expression, which indicates the spread of the wave packet.

\section{ACTION VARIABLE}

It is now of interest to calculate the action variable $J$ for the bound state
orbits in the harmonic oscillator problem considered above.  If we define
$J\equiv \oint m\dot{x}\; dx$, where $m\dot{x}$ is given by Eq.
(\ref{eq:xdot2}), it can be verified that $J=nh$ ($n=0,1,2,3..$) for those
particles whose complex paths enclose all the singular points of the integrand.
In the general case, we shall write

$$ J= \frac{\hbar}{i} \oint \frac{\partial \Psi /\partial x}{\Psi }\, dx $$
 We
may note that $\int \frac{\partial \Psi /\partial x}{\Psi }\, dx = \ln{\Psi } +
\mbox{constant}$ has a branch cut whenever $\Psi$ goes from positive to zero
through negative.  In the harmonic oscillator problem, the integral reduces to
$$ J= \frac { \hbar}{i} \oint \left[ -\alpha ^{2} x+ \frac {H_n^{\prime}(\alpha
x)}{H_n (\alpha x)}\right]dx = \frac { \hbar}{i} \oint \frac
{H_n^{\prime}(\alpha x)}{H_n (\alpha x)} dx $$
 where the prime denotes
differentiation with respect to $x$.  It is easily verified that for $n=0$, $J$
is always zero.  For $n=1$, $ J= \frac{\hbar}{i} \oint \frac{1}{x }\, dx = h$
when the integration is performed over complex paths enclosing the pole $x=0$.
From Fig.  \ref{fig:shm1}, it is seen that this happens for the larger nests
with $A>1$ whereas for the subnests with $A<1$, $J=0$.  (In fact, the subnests
appear to be physically nonviable single particle solutions.)  Similarly for
$n=2$, $J=2h$ when the complex path encloses the poles at $x=\pm \frac
{1}{\sqrt{2}\alpha }$ and so on.  In all cases, $J$ vanishes for the subnests.

Comparing the value of the
action variable for the
present case with that of dBB and FFM reveals the fact that the new scheme is
quite different from the other representations.  In the dBB, $J=0$ by precept.
FFM has shown that the $J$-quantization by Milne \cite{milne} can be generalized
to allow microstates.  The integrand, the location of poles and the branch cuts
in the present case differ with that of Milne.  Also note that in Milne and FFM,
$J=nh$ for $n=1,2,3,..$ whereas in the new scheme $J=nh$, $n=0,1,2,3,..$ only
for the larger nests and $J=0$ otherwise.

  Having
obtained the value of the action variable, one may check whether $\partial
J/\partial E= \tau$ is the orbital period in this
scheme.  For the harmonic oscillator, this is true since $E$ is linear in $n$.
Whether $\tau$ is the orbital period for the general potential is an interesting
open question.  Following the arguments in \cite{goldstein}, one can see that in
spite of the complexification involved, $\tau\equiv 1/\nu$ is the orbital period
if the equation of motion for $w\equiv \partial \hat{W}/\partial E$ is given by
$\dot{w}=\partial H(J)/\partial J$.  We need to solve more examples to verify
this.

Under these circumstances, one may conclude that the
issue of action variable is not satisfactorily resolved for the new scheme.

\section{DISCUSSION}

In summary, we find that the dBB identification $\Psi =Re^{iS/\hbar}$, which
itself is quite arbitrary, does not help to utilize all the information
contained in the wave function while solving the equation of motion
$m\dot{x}=\partial S/\partial x $.  In the present formulation we use a quantum
Hamilton-Jacobi equation (obtained by substituting $\Psi = {\cal N}
e^{i\hat{S}/\hbar}$ in the Schrodinger equation), which is closer to the
classical one.  This formalism is already used by some authors (for example, see
\cite{brown}) for other purposes.  Employing the equation of state $m\dot{x} =
\partial \hat{S}/\partial x$ brings the motion to a complex $x$-plane.  The
physical coordinate or trajectory of the particle is postulated to be the real
part of $x(t)$.  Solving the equation of motion in some bound state problems
shows why the dBB scheme gives zero velocity for particles in these cases; the
complex paths cross the real axis parallel to the imaginary one, so that
$\dot{x}_r=0$ at these points.  The positive results we obtained for the
harmonic oscillator and potential step problems themselves are indicative of the
deep insight obtainable in such problems by the use of the modified dBB scheme.

It shall be noted that some of the differences between FFM and dBB trajectories
are still there between FFM and modified dBB trajectories.  But as claimed in
the paper, modified dBB has many advantages over dBB trajectories.  An outline
of such differences between FFM, dBB and modified dBB trajectories can be made
as follows:

									       (i)
dBB and modified dBB need probability to obtain expectation values of observable
physical quantities.  But FFM trajectories do not involve probability.

	(ii) dBB and modified dBB have similar equations of
motion.  FFM uses a different equation of motion.

									       (iii)
dBB asserts that the possible trajectories for a particular particle should not
cross.  But FFM and modified dBB trajectories can
cross.

									       (iv)
For bound states, the particle in the dBB scheme does not move.  But in FFM and
modified dBB approaches, the particles in bound states have well-defined
trajectories.

									       (v)
The ultimate test of these trajectories lies in experiments involving time
sequence of single systems (which can be carried out
nowadays), rather than those involving large number of particles.  FFM has
already proposed some experiments which involve predictions different from
standard quantum mechanics.  In that sense, FFM is a testable theory.  The
present modified dBB formulation is at a disadvantage that as yet it does not
provide a physical example where such predictions arise.  An important avenue to
search for in this alternative approach is a solution to this shortcoming.

\medskip

									       A
theoretical understanding of the correctness of the equation of motion used
here, in comparison with the equation of motion used in
FFM is highly desirable.  This is particularly so since we cannot expect both
theories to be true while they differ.  This comparison can perhaps be made by
analysing the time concept (Floydian time or some standard time) and could be
clarified therefrom.  Also this important question is not addressed in this
paper.

									       The
prescription of $\dot{x}$ by Eq.  (\ref{eq:xdot2}) is reminiscent of
Schrodinger's seminal work \cite{jammer} on the wave function.  Schrodinger
attempted to derive a wave equation by accepting the classical
Hamilton-Jacobi equation as a phenomenological equation and by replacing
$\partial W/\partial x$ in this equation with $(K/\Psi )(\partial \Psi /\partial
x)$ where $K$ is a constant.  The present work can be considered to be a
reversal of Schrodinger's process, in which the attempt is to obtain an equation
of motion for the particle by accepting the Schrodinger wave equation as a
phenomenological equation and by equating $(\hbar /i \Psi ) (\partial \Psi
/\partial x)$ with $m\dot{x}$.  In this sense, it is closer to Schrodinger's
than to dBB.

 Generalization
of the formalism to more than one dimension is not attempted in this letter.
Its application to those other physical problems of interest shall be addressed
in future publications.

\medskip

\begin{center} 
{\bf Acknowledgements} 
\end{center}

 Valuable
comments made by Drs.  Edward R.  Floyd, A.  Bouda, M.  Matone, R.  Brown and R.
Carroll are acknowledged with thanks.  It is a pleasure to thank Professor K.
Babu Joseph, Sajith and Satheesh for encouragement and discussion.

\end{document}